\documentclass[pra,twocolumn,amsmath,amssymb,showpacs]{revtex4-1}

\usepackage[T1]{fontenc} 
\usepackage{bm}
\usepackage{graphicx}
\usepackage[bookmarks=false]{hyperref}
\usepackage{color}

\newcommand{\bra}[1]{\langle #1|}
\newcommand{\ket}[1]{|#1\rangle}
\newcommand{\braket}[1]{\langle #1 \rangle}

\def\dd{\mathrm{d}}
\def\ee{\mathrm{e}}
\def\ii{\mathrm{i}}

\def\Re{\mathrm{Re}}

\def\cc{\mathrm{c.c.}}
\def\ddt{(\partial/\partial t)}

\def\diez{\varepsilon_0}

\def\dens{\rho}

\def\ocu{Z_{\text{p}}}
\def\Zth{Z_{\text{th}}}

\def\gammaL{\gamma_{\parallel}}

\def\gammapure{\gamma_{\text{pure}}}
\def\gammatotal{\gamma_{\text{total}}}

\def\dm{d}
\def\wa{\omega_{\text{a}}}
\def\wc{\omega_{\text{c}}}

\def\rabi{g}
\def\rabia{g'}

\def\oH{\hat{H}}

\def\oHCC{\hat{H}_{\text{Coulomb}}^{\text{cavity}}}

\def\oHPC{\hat{H}_{\text{dipole}}^{\text{cavity}}}

\def\oU{\hat{U}}

\def\oa{\hat{a}}
\def\oad{\hat{a}^{\dagger}}
\def\osigma{\hat{\sigma}}

\def\oS{\hat{S}}

\def\oA{\hat{A}}
\def\oP{\hat{P}}

\def\vr{\bm{r}}

\def\AAC{\bar{\mathcal{A}}}
\def\BBC{\bar{\varPi}}
\def\XXC{\bar{X}}
\def\YYC{\bar{Y}}
\def\ZZC{\bar{Z}}

\def\AAP{\tilde{\mathcal{A}}}
\def\BBP{\tilde{\varPi}}
\def\XXP{\tilde{X}}
\def\YYP{\tilde{Y}}
\def\ZZP{\tilde{Z}}

\def\AA{\mathcal{A}}
\def\BB{\varPi}
\def\XX{X}
\def\YY{Y}
\def\ZZ{Z}

\def\aaC{\bar{\alpha}}
\def\bbC{\bar{p}}
\def\xxC{\bar{x}}
\def\yyC{\bar{y}}
\def\zzC{\bar{z}}

\def\aaP{\tilde{\alpha}}
\def\bbP{\tilde{p}}
\def\xxP{\tilde{x}}
\def\yyP{\tilde{y}}
\def\zzP{\tilde{z}}

\def\aa{\alpha}
\def\bb{p}
\def\xx{x}
\def\yy{y}
\def\zz{z}

\def\oHz{\hat{H}_0}
\def\oHSEC{\hat{H}_{\text{SEC}}}

\def\os{\hat{s}}
\def\osd{\hat{s}^{\dagger}}
\def\of{\hat{f}}
\def\ofd{\hat{f}^{\dagger}}

\def\oF{\hat{F}}
\def\oFd{\hat{F}^{\dagger}}

\def\oO{\hat{O}}
\def\orho{\hat{\rho}}

\begin{document}


\title{Laser under ultrastrong electromagnetic interaction with matter}

\author{Motoaki Bamba}
\altaffiliation{Present address:
Department of Materials Engineering Science, Osaka University,
1-3 Machikaneyama, Toyonaka, Osaka 560-8531, Japan\\
E-mail: bamba@qi.mp.es.osaka-u.ac.jp}
\affiliation{Department of Physics, Osaka University, 1-1 Machikaneyama, Toyonaka, Osaka 560-0043, Japan}
\author{Tetsuo Ogawa}
\affiliation{Department of Physics, Osaka University, 1-1 Machikaneyama, Toyonaka, Osaka 560-0043, Japan}

\date{\today}

\begin{abstract}
The conventional picture of the light amplification by stimulated emission of radiation (laser)
is broken under the ultrastrong interaction
between the electromagnetic fields and matter,
and distinct dynamics of the electric field and of the magnetic one
make the ``laser'' qualitatively different from the conventional laser,
which has been described simply without the distinction.
The ``laser'' in the ultrastrong regime
can show a rich variety of behaviors
with spontaneous appearance of coherence.
We found that the ``laser'' generally accompanies odd-order harmonics of the electromagnetic fields
both inside and outside the cavity
and a synchronization with an oscillation of atomic population.
A bistability is also demonstrated
in a simple model under two-level and single-mode approximations.
\end{abstract}




\maketitle
The light and microwave amplification by stimulated emission of radiation
(laser and maser)
were realized in 1960 \cite{Maiman1960N} and 1958 \cite{Schawlow1958PR},
respectively.
Although the fundamental theory for them is established
up to the quantum fluctuations of light and microwave
\cite{Haken1970,Haken1985,Scully1997,gardiner04},
the discussion is performed basically
under the rotating-wave approximation (RWA) on the interaction
between the electromagnetic fields and matter.
Under the RWA,
the total number of photons and atomic excitations is conserved during the interaction,
and it has enabled the simple picture based on the photons and excitations.
However, the RWA fails in the ultrastrong interaction regime,
that shows vacuum Rabi splitting comparable to
transition frequency of the atomic excitation \cite{Ciuti2005PRB}
and can now be realized in a variety of systems experimentally
\cite{Gunter2009N,Anappara2009PRB,Todorov2009PRL,Todorov2010PRL,
Niemczyk2010NP,Fedorov2010PRL,Forn-Diaz2010PRL,
Schwartz2011PRL,Porer2012PRB,Scalari2012S}.
In this regime,
we will see that dynamics of the electric field and of the magnetic one
should be discussed distinctively due to the lack of the RWA,
and we can no longer describe the laser
by the stimulated emission of radiation
without the distinction.
Resulting from this additional degree of freedom
originating from the distinction,
the ``laser'' and ``maser'' in the ultrastrong regime
are expected to show essential differences from the conventional laser.

The additional degree of freedom appears
in the equations of the ``laser'' (including the meaning of ``maser'' in the followings),
and we will see that the ``laser'' solutions must have multiple harmonics
for satisfying them.
Due to the complicated equations with the multi-harmonic expansion,
we can find a rich variety of ``laser'' solutions
that were hidden under the RWA in the conventional laser theory.
In other words, the recovery of the original distinction
of the electromagnetic fields in the ultrastrong regime
brings out the multiple harmonics and the rich solutions.
This is the conclusion of this paper,
and we will show, as a demonstration,
that bistable ``laser'' solutions are obtained
even by a basic calculation.

In order to highlight the qualitative and essential aspects of the ``laser''
in the ultrastrong regime,
we suppose an ensemble of all identical two-level atoms
interacting with a single mode in a cavity of the electromagnetic fields,
which has been considered to catch the basic properties
of the conventional laser \cite{Haken1970,Haken1985,Scully1997,gardiner04}.
This simple system keeps the generality of the conventional laser theory,
although it is known \cite{Todorov2010PRL,Todorov2012PRB,Todorov2014PRB}
that the finite-level and finite-mode approximations in the ultrastrong regime
are not suitable for pursuing the quantitative reliability,
which is obtained only by specifying a particular system of interest.
Instead, the calculations in this paper are performed in two typical gauges:
the Coulomb gauge and the electric-dipole one
\cite{cohen-tannoudji89,Keeling2007JPCM,Vukics2014PRL,Bamba2014SPT}.
The qualitative and general properties
should be obtained independently of the gauge choice,
which however gives us quantitatively different results.

Under the two-level, single-mode, and long-wavelength approximations,
the system Hamiltonians are obtained in the Coulomb and electric-dipole gauges, respectively, as
\cite{Bamba2014SPT,Note1}
\begin{align} \label{eq:oHCC} 
\oHCC
& = \hbar\wc\oad\oa
+ \sum_{\lambda=1}^N \frac{\hbar\wa}{2} \osigma_{\lambda}^z
+ \frac{\rabi\hbar\wa}{\sqrt{N}} ( \oa + \oad )
  \sum_{\lambda=1}^N \osigma_{\lambda}^y
\nonumber \\ & \quad
+ \rabi^2\hbar\wa \left( \oa + \oad \right)^2, \\
\label{eq:oHPC} 
\oHPC
& = \hbar\wc\oad\oa
  + \sum_{\lambda=1}^N\frac{\hbar\wa}{2}\osigma_{\lambda}^z
- \frac{\ii\rabi\hbar\wc}{\sqrt{N}} ( \oa - \oad )
  \sum_{\lambda=1}^N \osigma_{\lambda}^x
\nonumber \\ & \quad
+ \frac{\rabi^2\hbar\wc}{N}
  \sum_{\lambda=1}^N\sum_{\lambda'=1}^N
  \osigma_{\lambda}^x \osigma_{\lambda'}^x.
\end{align}
Here, $\wc$ is the frequency of the cavity mode,
and $\wa$ is the atomic transition frequency.
$\oa$ is the annihilation operator of a photon
in the cavity mode and satisfies $[\oa,\oad]=1$,
while the photon does not provide a good picture in the ultrastrong regime.
$N$ is the number of atoms,
and $\osigma_{\lambda}^{x,y,z}$ is the Pauli matrix representing
the $\lambda$-th atom.
$\rabi$ is a non-dimensional interaction strength:
\begin{equation}
\rabi = \sqrt{\frac{\dens|\dm|^2}{2\diez\hbar\wc}},\quad
\rabia = \rabi\sqrt{\frac{\wc}{\wa}} = \sqrt{\frac{\dens|\dm|^2}{2\diez\hbar\wa}},
\end{equation}
where $\dm$ is the transition dipole moment of the atomic transition,
$\diez$ is the vacuum permittivity,
and $\dens$ is the density of atoms.
In this paper, we define the ultrastrong regime
as $\rabia\gtrsim1$, which is determined only by the atomic parameters.

Under the RWA, the counter-rotating terms
$\oa\osigma_{\lambda}$ and $\oad\osigma^{\dagger}_{\lambda}$ are neglected
in the interaction Hamiltonians
($\osigma_{\lambda}=(\osigma^x_{\lambda}-\ii\osigma^y_{\lambda})/2$
is lowering operator of $\lambda$-th atom),
and the total number of photons and atomic excitations is conserved
in the remaining terms $\oad\osigma_{\lambda}$ and $\osigma^{\dagger}_{\lambda}\oa$.
Thanks to this simplification, in the conventional laser theory,
we need to consider only the following three variables:
light amplitude $\braket{\oa}/\sqrt{N}$,
atomic one $\sum_{\lambda}\braket{\osigma_{\lambda}}/N$,
and atomic population $Z=\sum_{\lambda}\braket{\osigma_{\lambda}^{z}}/2N$
\cite{Haken1970,Haken1985,Scully1997,gardiner04}.
However, in the ultrastrong regime,
this simple picture is no longer appropriate
($\oa$ and $\osigma_{\lambda}$ no longer associate positive-frequency components
or lowering operators of the system)
due to the lack of the RWA.
Instead, we need to consider the following five variables distinctively:
the non-dimensional vector potential $\AA=\braket{\oa+\oad}/\sqrt{N}$,
electric field (or displacement) $\BB=\ii\braket{\oa-\oad}/\sqrt{N}$,
atomic polarization $\XX=\sum_{\lambda}\braket{\osigma_{\lambda}^{x}}/2N$,
current $\YY=\sum_{\lambda}\braket{\osigma_{\lambda}^{y}}/2N$,
and population $\ZZ$.
Thanks to this distinction or the additional degrees of freedom,
we can find unconventional solutions
of complicated ``laser'' equations in the ultrastrong regime.

The inverted population is inevitable for the laser in our system even in the ultrastrong regime,
and it corresponds to $\ZZ>0$, while $|\ZZ|\leq1/2$.
Here, we introduce heat baths for pumping the atoms incoherently
and also a bath for dissipation of the electromagnetic fields,
whose couplings are mediated by $\osigma_{\lambda}^x$ and $(\oa+\oad)$,
respectively, for suppressing the gauge-dependence \cite{Note10}.
Without the electromagnetic interaction with matter,
each atom is incoherently pumped to $\ocu$
with a rate $\gammaL$ \cite{gardiner04},
and the electromagnetic fields decay with a rate $\kappa$.
We also consider baths for pure dephasing of atomic amplitudes $\XX$ and $\YY$
(including influence of broadening of atomic transition frequencies),
and they are mediated by $\osigma_{\lambda}^z$
with a rate $\gammapure$.
These dissipation rates are supposed to be frequency-independent for simplicity.

In the presence of these system-environment couplings,
we derived quantum stochastic differential equations \cite{gardiner04}
based on the positive/negative frequency components \cite{Bamba2014SEC}
(see also the supplemental material \cite{Note2}).
For large enough $N\gg1$ \cite{gardiner04},
macroscopic ``laser'' equations factorized by the above five variables
are obtained
in the Coulomb and electric-dipole gauges, respectively, as
\begin{subequations} \label{eq:laserC} 
\begin{align}
\ddt\AAC & = - \wc \BBC, \label{eq:AAC} \\
\ddt\BBC & = (\wc+4g^2\wa) \AAC - \ii\kappa[\AAC^{(+)}-\AAC^{(-)}]+ 4\rabi\wa\YYC, \\
\ddt\XXC & = -\gamma_x\XXC - \wa\YYC + 2\rabi\wa\AAC\ZZC, \label{eq:XXC} \\
\ddt\YYC & = \wa \XXC - \gamma_y\YYC, \\
\ddt\ZZC & = -\gamma_z(\ZZC-\ocu) - 2\rabi\wa\AAC\XXC,
\label{eq:ZZC} 
\end{align}
\end{subequations}
\begin{subequations} \label{eq:laserP} 
\begin{align}
\ddt\AAP
& = - \wc \BBP + 4\rabi\wc \XXP, \\
\ddt\BBP
& = \wc \AAP -\ii\kappa [\AAP^{(+)}-\AAP^{(-)}], \\
\ddt\XXP
& = - \gamma_x\XXP - \wa \YYP, \\
\ddt\YYP
& = (\wa - 8\rabi^2\wc\ZZP) \XXP -\gamma_y\YYP + 2\rabi\wc\BBP\ZZP, \label{eq:YYP} \\
\ddt\ZZP
& = -\gamma_z(\ZZP-\ocu)
  - 2\rabi\wc \BBP \YYP + 8\rabi^2\wc \XXP \YYP.
\label{eq:ZZP} 
\end{align}
\end{subequations}
Here,
the variables in the Coulomb gauge are distinguished by bar as $\AAC$,
and those in the electric-dipole gauge are distinguished by tilde as $\AAP$.
The superscript $(\pm)$ means the positive/negative frequency component
of the variables,
based on which we get asymmetric dissipation rates:
$\kappa$ only for $\varPi$,
$\gamma_x=\gammapure$, $\gamma_y=\gammaL + \gammapure$,
and $\gamma_z=\gammaL$ \cite{Note3}.

In the conventional laser, the electromagnetic and atomic amplitudes
($\AA,\BB,\XX$, and $\YY$) oscillate
with a frequency $\varOmega$ (determined for satisfying the laser equations),
and the atomic population $\ZZ$ is constant in time.
In contrast, in the ultrastrong regime,
since the RWA cannot be applied to $2\rabi\wa\AAC(t)\XXC(t)$ in Eqs.~\eqref{eq:ZZC}
and also to $2\rabi\wc\BBP(t)\YYP(t)$ in Eq.~\eqref{eq:ZZP},
the atomic population $\ZZC(t)$ and $\ZZP(t)$ are driven
not only by the time-constant components
but also by $2\varOmega$ ones,
when the amplitudes oscillate with a fundamental frequency $\varOmega$.
Further, $\XXC$ and $\YYP$ are driven by $3\varOmega$ components through
$2\rabi\wa\AAC(t)\ZZC(t)$ in Eq.~\eqref{eq:XXC}
and $2\rabi\wc\BBP(t)\ZZP(t)$ in Eq.~\eqref{eq:YYP},
respectively.
Therefore, the electromagnetic and atomic amplitudes in general
oscillate with odd-order harmonics
$\varOmega$, $3\varOmega$, $5\varOmega$, \dots,
and the atomic population oscillates
with even-order harmonics $0\varOmega$, $2\varOmega$, $4\varOmega$, \dots.
Since the cavity loss is mediated by $\AA$,
the dynamics of the frequency components $\AA^{(\pm)}(t)$ reflect those of the output from the cavity
through the input-output relation \cite{gardiner04,Ridolfo2012PRL,Bamba2014SEC}.
Then, the output also oscillates with the odd-order harmonics.
These multiple harmonics and synchronization with the atomic population
are obtained in general.
They are a part of the qualitative differences from the conventional laser
and are obtained independently of the gauge choice.

For pursuing the simplicity,
we consider only up to the third-order harmonic,
which is sufficient for finding bistable ``laser'' solutions.
The electromagnetic and atomic amplitudes are expanded as
$\AA(t)=\aa_{1}\ee^{-\ii\varOmega t}+\aa_{3}\ee^{-\ii3\varOmega t}+\cc$,
$\BB(t)=\bb_{1}\ee^{-\ii\varOmega t}+\bb_{3}\ee^{-\ii3\varOmega t}+\cc$
(also for $\XX$ and $\YY$),
and the atomic population as $\ZZ(t)=\ZZ_0+(\zz_2\ee^{-\ii2\varOmega t}+\cc)$.
Neglecting highly oscillating terms (RWA in oscillation basis),
non-trivial oscillating steady states are obtained
from Eqs.~\eqref{eq:laserC} and \eqref{eq:laserP} \cite{Note2},
and they correspond to the ``laser'' states.

\begin{figure}[tbp]
\includegraphics[width=\linewidth]{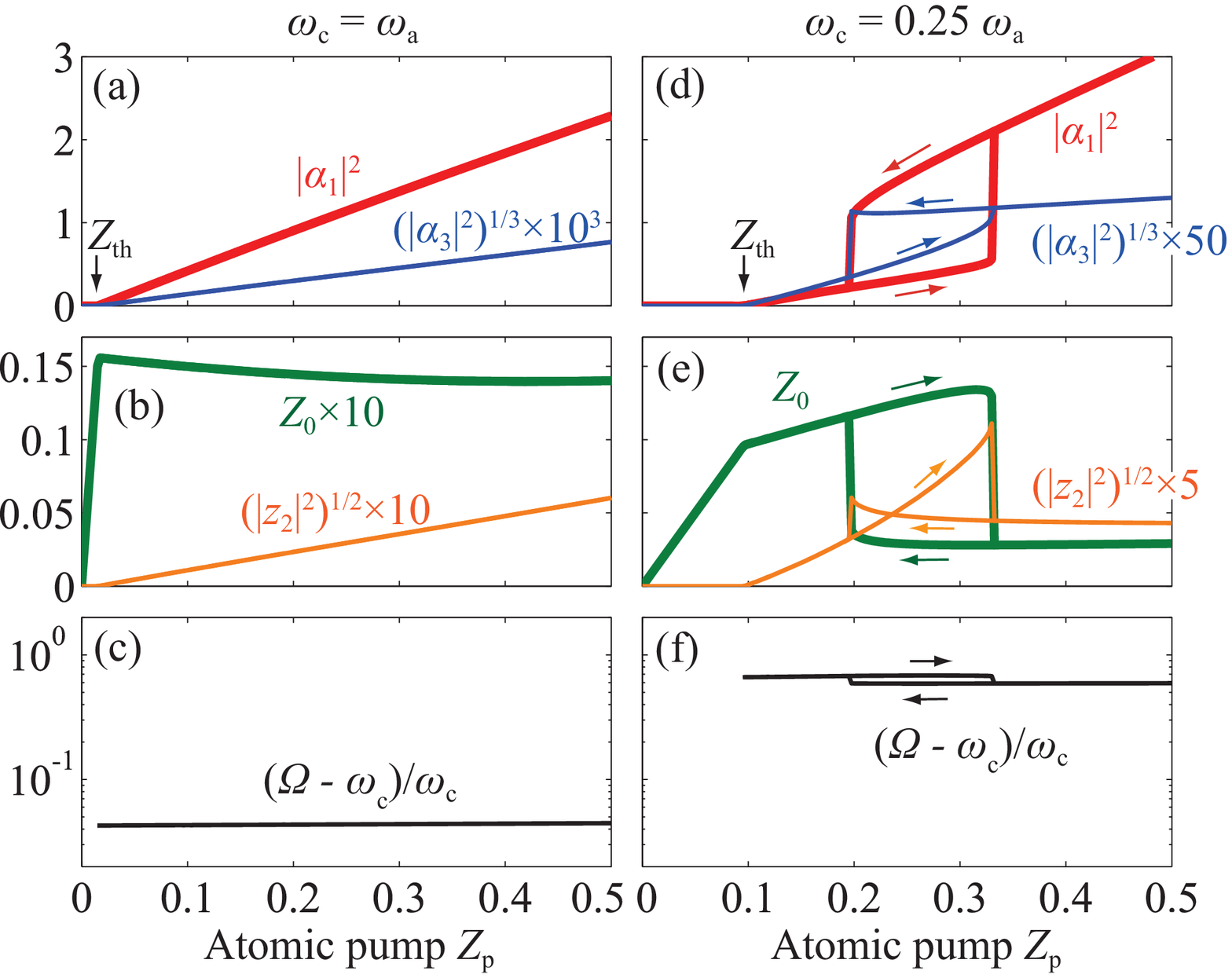}
\caption{``Laser'' solutions are calculated
with increasing and decreasing the atomic pump $\ocu$ in the Coulomb gauge.
The bare cavity frequencies are (a-c) $\wc=\wa$ and (d-f) $\wc=0.25\wa$.
(a,d) Intensities of fundamental component $\aa_1$ and $3\varOmega$ one $\aa_3$
of non-dimensional vector potential $\AA = \braket{\oa+\oad}/\sqrt{N}$,
(b,e) time-constant component $\ZZ_0$ and $2\varOmega$ one $\zz_2$
of the atomic population,
and (c,f) $(\varOmega-\wc)/\wc$ for fundamental oscillation frequency $\varOmega$
are plotted.
Below threshold $\ocu<\Zth$, we get $\ZZ_0=\ocu$ and zero oscillating components.
Above threshold, we get a linear increase of $|\aa_1|^2$,
$|\zz_2|\propto|\aa_1|^2$, and $|\aa_3|=|\aa_1|^3$ for $\wc=\wa$.
A bistability appears for $\wc=0.25\wa$.
The arrows in the figures represent the solutions
with increasing and decreasing $\ocu$.
Parameters: $\rabia=0.15$, $\gammaL=0.05\wa$, $\gammapure=0.1\wa$,
and $\kappa=0.01\wa$.}
\label{fig:1}
\end{figure}
In Fig.~\ref{fig:1}, we plot the ``laser'' solutions
in the Coulomb gauge versus the atomic pump $\ocu$.
The interaction strength is supposed as $\rabia=0.15$,
which is relevant as reported for the organic molecules \cite{Schwartz2011PRL}.
We suppose the atomic dissipation rates as
$\gammaL=0.05\wa$ and $\gammapure=0.1\wa$
by considering currently available samples.
The supposed cavity loss rate $\kappa=0.01\wa$ is lower than the one in Ref.~\cite{Schwartz2011PRL},
but much lower rate is available by distributed Bragg reflectors
\cite{Koschorreck2005APL,Han2013APE,Akselrod2014PRB}.
In Fig.~\ref{fig:1}(a-c),
the bare cavity frequency is equal to the atomic one as $\wc=\wa$.
Below the threshold $\Zth=1.56\times10^{-2}$,
the atomic population simply increases with obeying $\ZZ=\ZZ_0=\ocu$,
and the oscillation components are zero.
Above the threshold, 
the intensity of the fundamental oscillation component
increases linearly as $|\aa_1|^2\propto(\ocu-\Zth)$
as seen in Fig.~\ref{fig:1}(a),
and the time-constant component $\ZZ_0$ of the atomic population
is almost unchanged after the threshold
as in Fig.~\ref{fig:1}(b).
These are similar as the conventional laser.

The difference is the appearance of the multiple harmonics ($\zz_2$ and $\aa_3$),
which are generally obtained
even though higher cavity modes with $2\wc,3\wc,\ldots$
are not considered \cite{Note4}.
As seen in Figs.~\ref{fig:1}(a,b),
The multiple harmonics increase as
$|\zz_2|^2\propto(\ocu-\Zth)^2$ and $|\aa_3|^2\propto(\ocu-\Zth)^3$
as the third-order nonlinear effect (perturbation)
of the fundamental component $\aa_1$.

\begin{figure}[tbp]
\includegraphics[width=\linewidth]{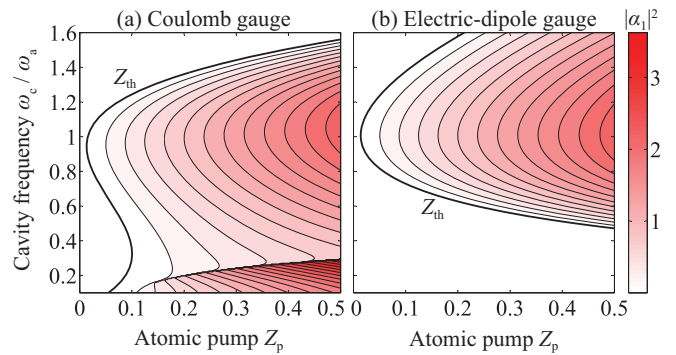}
\caption{Maps of ``laser'' solutions in (a) Coulomb gauge and (b) electric-dipole one.
The intensity $|\aa_1|^2$ of the fundamental component is calculated
by increasing $\ocu$ for a fixed $\wc$, which is also changed in the vertical axis.
The bold curves indicate the threshold $\Zth$.
Unconventional solutions appear for low cavity frequency $\wc<0.29\wa$
in the Coulomb gauge.
Although we find only the conventional solutions in the electric-dipole gauge
under the current parameters,
the unconventional solutions appear for larger $\rabia$ or lower $\kappa$ \cite{Note2}.
This quantitative difference is caused by the two-level and single-mode approximations
used in the calculation.
Parameters: $\rabia=0.15$, $\gammaL=0.05\wa$, $\gammapure=0.1\wa$,
and $\kappa=0.01\wa$.}
\label{fig:2}
\end{figure}
In contrast, in Figs.~\ref{fig:1}(d-f),
where the cavity frequency is far below the atomic resonance as $\wc=0.25\wa$,
we can find a bistable behavior above the threshold $\Zth=9.62\times10^{-2}$
\cite{Note5}.
In Fig.~\ref{fig:2}(a), 
we plot the intensity $|\aa_1|^2$ of the fundamental component
calculated in the Coulomb gauge
by increasing $\ocu$ for a fixed $\wc$,
which is also changed in the vertical axis.
When $\wc$ is around the atomic frequency $\wa$,
the threshold $\Zth$ is minimized and $|\aa_1|^2$ is locally maximized.
The bistability appears for low cavity frequency $0.15\wa<\wc<0.29\wa$.
For $\wc<0.15\wa$, we do not find a clear jump,
and the solutions change continuously (but drastically) with the increase of $\ocu$.

The appearance of the bistability can be understood
simply by the fact that
the third harmonic becomes close to the atomic resonance
[$3\varOmega\sim1.2\wa$ in Fig.~\ref{fig:1}(d-f)].
Comparing with Fig.~\ref{fig:1}(a) ($3\varOmega\sim3.1\wa$),
the $3\varOmega$ component $\aa_3$ is significantly enhanced in Fig.~\ref{fig:1}(d),
while the fundamental one $\aa_1$ is in the same order.
Thanks to the relatively large amplitudes of the multiple harmonics,
we can find unconventional solutions for the complicated nonlinear equations
\eqref{eq:laserC} and \eqref{eq:laserP}
with the five variables (or more in the multi-harmonic expansion).
This is the reason why the bistability appears for the low cavity frequency
in Fig.~\ref{fig:2}(a).
For enlarging the $\wc$-range showing the ``laser'' down to such a low frequency,
strong $\rabia$, low $\kappa$, and high $\gammatotal=\gammaL+\gammapure$ are desired
(for $\kappa<\gammatotal$) as expected from the conventional laser theory
\cite{Haken1970,Haken1985,Scully1997,gardiner04}.
Further, low pure-dephasing ratio $\gammapure/\gammatotal$
($=2/3$ in present calculation)
is advantageous to enhancing the multi-harmonic amplitudes,
and then the bistability appears more clearly.
This is checked numerically in supplemental material \cite{Note2}.

The signature of the electromagnetic distinction
appears particularly in the bistable region.
In the resonant case ($\wc=\wa$),
the interaction is suppressed effectively
through $2\rabi\wa\AAC\ZZC$ in Eq.~\eqref{eq:XXC}
and $2\rabi\wc\BBP\ZZP$ in Eq.~\eqref{eq:YYP}
by the negligible atomic population $\ZZ\sim0.01$
shown in Fig.~\ref{fig:1}(b).
Then, the ``laser'' is reduced approximately to the conventional one,
and the multiple harmonics appear perturbatively.
In this sense, the bistability in Fig.~\ref{fig:1}(d-f) is
correlated strongly to the electromagnetic distinction, because the interaction
is not significantly suppressed
by the atomic population $\ZZ\sim0.1$ as seen in Fig.~\ref{fig:1}(e).
The signature of the distinction is also found in Figs.~\ref{fig:1}(c,f)
showing $(\varOmega-\wc)/\wc=|\bb_1/\aa_1|-1$,
i.e., the amplitude difference between the non-dimensional
vector potential and the electric field
(this relation is obtained from Eq.~\eqref{eq:AAC} in the Coulomb gauge).
The negligible difference $|\bb_1/\aa_1|-1\ll1$ in Fig.~\ref{fig:1}(c)
corresponds approximately to the conventional laser, 
in which the photons are well defined as $|\aa_1|=|\bb_1|$.
In contrast, in the bistable case,
the relatively large $|\bb_1/\aa_1|-1$ in Fig.~\ref{fig:1}(f) indicates that
the electric field $\BB$ and the magnetic one
(or vector potential $\AA$) shows clearly distinct dynamics
including the multiple harmonics.
This is certainly what we initially expected in the ultrastrong regime,
and the bistability originates from this distinction \cite{Note6}.

\begin{figure}[tbp]
\includegraphics[width=\linewidth]{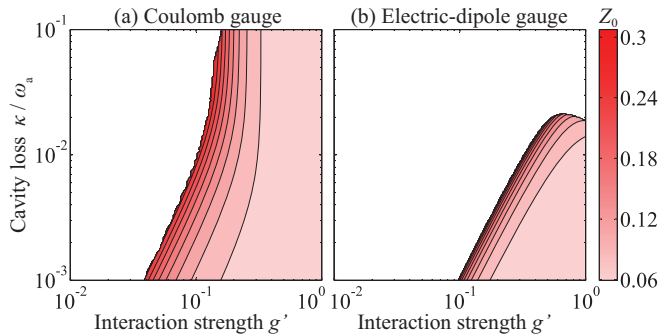}
\caption{The region of parameters $\rabia$ and $\kappa/\wa$ showing the bistability
is plotted in (a) Coulomb gauge and (b) electric-dipole one.
The color indicates $\ZZ_0$ (a measure of electromagnetic distinction)
for the highest bare cavity frequency
that shows the bistability at $\ocu=0.5$.
In both gauges, the bistability appears
for strong enough interaction $\rabia$
and low enough cavity loss $\kappa$.
Parameters: $\gammaL=0.05\wa$ and $\gammapure=0.1\wa$.}
\label{fig:3}
\end{figure}
In Fig.~\ref{fig:2}(b), the ``laser'' solutions calculated
in the electric-dipole gauge are plotted under the same parameters
as in Fig.~\ref{fig:2}(a).
Although the bistability does not appear in Fig.~\ref{fig:2}(b),
it is just the quantitative difference
caused by the two-level and single-mode approximations,
and we can obtain the bistability even in the electric-dipole gauge
for different parameters \cite{Note2}.
In Fig.~\ref{fig:3},
the parameter regions showing the bistability are plotted in the two gauges.
We plot $\ZZ_0$ in the steady state
for the highest bare cavity frequency that shows the bistability for $\ocu=0.5$
($\wc=0.29\wa$ in Fig.~\ref{fig:2}(a))
as functions of $\rabia$ and $\kappa$.
The bistability appears basically for the strong enough $\rabia$
and low enough $\kappa/\wa$,
both of which is necessary for obtaining the ``laser''
around $3\varOmega\sim\wa$ as discussed above.
The population $\ZZ_0$ is a measure of the electromagnetic distinction.
The bistability starts to appear with a large $\ZZ_0$,
which enhances the distinction of the electromagnetic fields,
and $\ZZ_0$ still keeps a certain value ($\sim0.06$ at current parameters)
even when the bistability is easily found for large enough $\rabia$.

The quantitative difference for the appearance of the bistability
mathematically originates from the following fact.
As seen in Eqs.~\eqref{eq:oHCC} and \eqref{eq:oHPC},
the interaction terms are proportional to $\rabi\wa$ and $\rabi\wc$
in the Coulomb and electric-dipole gauges, respectively.
Since the coefficient $\rabi\wa\propto1/\sqrt{\wc}$
is increased with the decrease of $\wc$ in the Coulomb gauge,
the ``laser'' solution is easily found for low $\wc$.
As the result, the bistability is easily found in the Coulomb gauge
compared to the electric-dipole one.
Although this quantitative gauge-dependence
is diminished if we consider all the atomic levels and the cavity modes
by specifying particular systems of interest
\cite{Todorov2012PRB,Todorov2014PRB,Bassani1977PRL},
the two-level approximation
is justified qualitatively if the two atomic levels are well separated
by more than $\rabi\wa$ or $\rabi\wc$
from the other levels energetically \cite{Note7}.
Since the higher cavity modes basically
enhances the amplitudes of the multiple harmonics,
the bistability (or multi-stability) is also expected
beyond the single-mode approximations \cite{Note8}.
Whereas the present calculation still has such quantitative problems,
the bistability is expected to appear as
another qualitative difference from the conventional laser.

We conclude that, in the ultrastrong regime,
the ``laser'' generally accompanies odd-order harmonics of the electromagnetic fields
both inside and outside the cavity
and the synchronization with the atomic population
oscillating with even-order harmonics.
Whereas we found a bistability by the calculation
up to the third harmonic under the two-level and single-mode approximations,
a richer variety of the ``laser'' solutions could be obtained thanks to
the recovery of the original distinction of the electromagnetic fields
in the ultrastrong regime,
which exposes the additional degrees of freedom
hidden by the RWA.

The properties of this ``laser'' are not fully elucidated in this paper,
and those to be investigated spread as extensively
as the conventional laser has been studied
from the viewpoints of quantum optics, nonlinear physics,
non-equilibrium physics, synergetics, etc.
For example, it is open to dispute
whether the ``laser'' output is
in a simple coherent state as the ideal conventional laser
\cite{Haken1970,Haken1985,Scully1997,gardiner04}
or a non-classical state can be directly
obtained thanks to the ultrastrong interaction
especially in the bistable regions \cite{Note9}.
Experimentally, the ``laser'' in the ultrastrong regime would be realized
by fabricating microcavities embedding organic (dye) molecules \cite{Schwartz2011PRL}
or superconducting circuits \cite{Astafiev2007N} with (a large number of) artificial atoms.
Quantum cascade lasers involving inter-subband transitions in semiconductor quantum wells
\cite{Gunter2009N,Anappara2009PRB,Todorov2009PRL,Todorov2010PRL,Porer2012PRB} are also promising,
while the present calculation does not exactly correspond to it.

\begin{acknowledgments}
M.~B.~thanks H.~Ishihara for discussion.
This work was funded by ImPACT Program of Council for Science, Technology and
Innovation (Cabinet Office, Government of Japan)
and by JSPS KAKENHI (Grant No.~26287087 and 24-632).
\end{acknowledgments}

\begin{widetext}
\appendix

In App.~\ref{sec:SDE}, we briefly explain the framework of stochastic differential equations,
by which the ``laser'' equations in the main text are derived,
and also show the Hamiltonian of system-environment couplings.
The detailed calculations of the oscillating steady states from the ``laser'' equations
are shown in Apps.~\ref{eq:steady_Coulomb} and \ref{eq:steady_dipole}
in the Coulomb and electric-dipole gauges, respectively.
In App.~\ref{eq:numerical_dipole},
the bistability of the ``laser'' in the electric-dipole gauge is demonstrated,
and the dependence on the atomic decoherence is also discussed with some numerical calculations.

\section{Stochastic differential equations} \label{sec:SDE}
As a general discussion,
we consider system-environment coupling expressed as
\begin{equation} \label{eq:oHSEC_sample} 
\oH_{\text{SEC}}^{\text{example}}
= \int_0^{\infty}\dd\omega\
  \left\{
    \hbar\omega \ofd(\omega)\of(\omega)
+ \ii\hbar\sqrt{\frac{\varGamma(\omega)}{2\pi}} \oS \left[ \ofd(\omega) - \of(\omega) \right]
\right\}.
\end{equation}
Here, $\oS$ is a Hermitian operator of system of interest
and $\of(\omega)$ is the annihilation operator of a boson in the environment.
$\varGamma(\omega)$ corresponds to the bare dissipation rate.
Further, the distribution in the environment is supposed as
\begin{subequations}
\begin{align}
\braket{\ofd(\omega)\of(\omega')} & = n(\omega)\delta(\omega-\omega'), \\
\braket{\of(\omega)\ofd(\omega')} & = [n(\omega)+1]\delta(\omega-\omega').
\end{align}
\end{subequations}
For deriving the master equations,
we usually perform the pre-trace rotating-wave approximation (RWA)
to the system-environment coupling as \cite{Bamba2014SEC}
\begin{equation}
\oH_{\text{SEC}}^{\text{example}}
\simeq \int_0^{\infty}\dd\omega\
  \left\{
    \hbar\omega \ofd(\omega)\of(\omega)
+ \ii\hbar\sqrt{\frac{\varGamma(\omega)}{2\pi}}
  \left[ \ofd(\omega)\os - \osd\of(\omega) \right]
\right\},
\end{equation}
where $\os$ is the positive-frequency component of $\oS$ defined
with eigenstates $\{\ket{\mu}\}$ of system Hamiltonian $\oHz$:
\begin{equation}
\os = \sum_{\mu,\nu>\mu} \ket{\mu}\braket{\mu|\oS|\nu}\bra{\nu}.
\end{equation}
However, even from Eq.~\eqref{eq:oHSEC_sample} and system Hamiltonian $\oHz$,
we can derive a master equation for density operator $\orho(t)$ in the Schr\"odinger picture
by performing partially the pre-trace RWA as \cite{Bamba2014SEC,gardiner04}
\begin{multline} \label{eq:master_partial_preRWA} 
\frac{\dd}{\dd t}\orho(t)
= \frac{1}{\ii\hbar}\left[ \orho(t), \oHz \right]
+ \sum_{\mu,\nu>\mu} \frac{\varGamma(\omega_{\nu,\mu})}{2} [n(\omega_{\nu,\mu})+1]\left\{
    \left[\oS, \orho(t)\{\os_{\mu,\nu}\}^{\dagger}\right]
  + \left[\os_{\mu,\nu}\orho(t), \oS\right]
  \right\} \\
+ \sum_{\mu,\nu>\mu} \frac{\varGamma(\omega_{\nu,\mu})}{2} n(\omega_{\nu,\mu}) \left\{
    \left[\oS, \orho(t)\os_{\mu,\nu}\right]
  + \left[\{\os_{\mu,\nu}\}^{\dagger}\orho(t), \oS\right]
  \right\},
\end{multline}
where $\omega_{\nu,\mu}$ is the frequency difference
from eigenstate $\ket{\mu}$ to $\ket{\nu}$ of $\oHz$,
and $\os_{\mu,\nu}$ is defined as
\begin{equation}
\os_{\mu,\nu} = \ket{\mu}\braket{\mu|\oS|\nu}\bra{\nu}.
\end{equation}
Even in the above derivation,
Eq.~\eqref{eq:master_partial_preRWA} guarantees the thermal equilibrium
$\orho = \ee^{-\oHz/k_BT}$ in the steady state under the Bose distribution
$n(\omega) = 1/(\ee^{\hbar\omega/k_BT} - 1)$
in the environment.

From Eq.~\eqref{eq:master_partial_preRWA}
for frequency-independent dissipation rate $\varGamma(\omega) = \varGamma$
and flat distribution $n(\omega) = n$,
the corresponding quantum stochastic differential equation (QSDE)
is obtained for system operator $\oO$ in Itoh's from as \cite{gardiner04}
\begin{multline} \label{eq:QSDE_partial_preRWA} 
\dd\oO = \frac{1}{\ii\hbar}\left[ \oO, \oHz \right]\dd t
+ \frac{\varGamma(n+1)}{2}\left\{\osd[\oO, \oS]+[\oS, \oO]\os\right\}\dd t \\
+ \frac{\varGamma n}{2}\left\{\os[\oO, \oS]+[\oS, \oO]\osd\right\}\dd t
- \sqrt{\varGamma}[ \oO, \oS ] \dd\oF(t)
+ \sqrt{\varGamma}\dd\oFd(t)[ \oO, \oS ],
\end{multline}
where the fluctuation operator satisfies
\begin{subequations}
\begin{align}
\dd\oF(t)^2 = \dd\oFd(t)^2 & = 0 \\
\dd\oFd(t)\dd\oF(t) & = n \dd t \\
\dd\oF(t)\dd\oFd(t) & = (n+1) \dd t
\end{align}
\end{subequations}
When we replace $\oS$ by $\os$ or $\osd$,
Eq.~\eqref{eq:QSDE_partial_preRWA} is certainly reduced to the QSDE
discussed in Ref.~\cite{gardiner04}.
Since Eq.~\eqref{eq:QSDE_partial_preRWA} only have the commutator
between $\oO$ and the original Hermitian operator $\oS$,
we do not need the knowledge of the eigenstates of $\oHz$,
which is generally hard to be calculated.

For the dissipation and incoherent pumping
(by heat bath with a negative temperature \cite{gardiner04}) of the system,
we consider the following system-environment coupling
\begin{multline} \label{eq:oHSEC} 
\oHSEC^{x}
= \int_0^{\infty}\dd\omega\
  \left\{
    \hbar\omega \ofd_A(\omega)\of_A(\omega)
+ \ii\hbar\sqrt{\frac{\kappa}{2\pi}}(\oa+\oad)\left[ \ofd_A(\omega) - \of_A(\omega) \right]
\right. \\
  + \hbar\omega \sum_{\lambda=1}^N \ofd_{X,\lambda}(\omega)\of_{X,\lambda}(\omega)
+ \ii\hbar\sqrt{\frac{\gammaL|\ocu|}{\pi}} \sum_{\lambda=1}^N
  \osigma_{\lambda}^{x}\left[ \ofd_{X,\lambda}(\omega) - \of_{X,\lambda}(\omega) \right] \\
\left.
  + \hbar\omega \sum_{\lambda=1}^N \ofd_{Z,\lambda}(\omega)\of_{Z,\lambda}(\omega)
+ \ii\hbar\sqrt{\frac{\gammapure}{4\pi}} \sum_{\lambda=1}^N
  \osigma_{\lambda}^z \left[ \ofd_{Z,\lambda}(\omega) - \of_{Z,\lambda}(\omega) \right]
\right\}
\end{multline}
The environment fields $\of_A(\omega)$, $\of_{X,\lambda}(\omega)$, and $\of_{Z,\lambda}(\omega)$
are not correlated with each other,
and their self-correlations are supposed as
\begin{subequations}
\begin{align}
\braket{\ofd_A(\omega)\of_A(\omega')} & = 0, \\
\braket{\of_A(\omega)\ofd_A(\omega')} & = \delta(\omega-\omega'),
\end{align}
\end{subequations}
\begin{subequations}
\begin{align}
\braket{\ofd_{X,\lambda}(\omega)\of_{X,\lambda'}(\omega')} & = \frac{1/2+\ocu}{2|\ocu|}\delta_{\lambda,\lambda'}\delta(\omega-\omega'), \\
\braket{\of_{X,\lambda}(\omega)\ofd_{X,\lambda'}(\omega')} & = \frac{1/2-\ocu}{2|\ocu|}\delta_{\lambda,\lambda'}\delta(\omega-\omega'),
\end{align}
\end{subequations}
\begin{subequations}
\begin{align}
\braket{\ofd_{Z,\lambda}(\omega)\of_{Z,\lambda'}(\omega')} & = 0, \\
\braket{\of_{Z,\lambda}(\omega)\ofd_{Z,\lambda'}(\omega')} & = \delta_{\lambda,\lambda'}\delta(\omega-\omega').
\end{align}
\end{subequations}
From these system-environment couplings,
the equations of motion of the c-number variables in the main text
are derived from Eq.~\eqref{eq:QSDE_partial_preRWA}.
For the derivation, we considered that the following term is approximately zero
\begin{equation} \label{eq:sxpsz-sxmsz=0} 
\braket{\osigma_{\lambda}^{x(+)}\osigma_{\lambda}^z + \osigma_{\lambda}^z\osigma_{\lambda}^{x(+)}}
- \braket{\osigma_{\lambda}^{x(-)}\osigma_{\lambda}^z + \osigma_{\lambda}^z\osigma_{\lambda}^{x(-)}}
\simeq 0.
\end{equation}
For deriving this, from the relations
$\osigma_{\lambda}^x = \osigma_{\lambda}^{x(+)} + \osigma_{\lambda}^{x(-)}$
and $\osigma_{\lambda}^z\osigma_{\lambda}^x = \ii\osigma_{\lambda}^y$,
we get the following relation
\begin{equation} \label{eq:sxpsz=-sxmsz} 
\braket{\osigma_{\lambda}^{x(+)}\osigma_{\lambda}^z + \osigma_{\lambda}^z\osigma_{\lambda}^{x(+)}}
= - \braket{\osigma_{\lambda}^{x(-)}\osigma_{\lambda}^z + \osigma_{\lambda}^z\osigma_{\lambda}^{x(-)}}.
\end{equation}
The left and right hand sides oscillate mainly with positive and negative frequencies, respectively.
Then, for satisfying Eq.~\eqref{eq:sxpsz=-sxmsz},
both brackets should be zero, and Eq.~\eqref{eq:sxpsz-sxmsz=0} can be neglected.
Further, we used the following approximation
\begin{equation} \label{eq:<xy>+<yx>}
- \sum_{\lambda=1}^N
\frac{\braket{\osigma_{\lambda}^{x(+)}\osigma_{\lambda}^y}
    + \braket{\osigma_{\lambda}^y\osigma_{\lambda}^{x(+)}}}{\ii N}
\simeq 1.
\end{equation}
The pumping level $\ocu$ is modulated by this factor
in the equations of motion.
When the RWA can be applied to the electromagnetic interaction with matter
in the photon-excitation basis,
we get $\osigma_{\lambda}^{x(+)} = \osigma_{\lambda}$
and $\osigma_{\lambda}^{y(+)} = \ii\osigma_{\lambda}$,
then we can get the above equality.
In the ultrastrong regime,
the equality is generally violated.
However, if we get $|\ZZ| \ll 1$, the strength of the interaction is effectively suppressed,
and $\osigma_{\lambda}^{x(+)} \simeq \osigma_{\lambda}$
and $\osigma_{\lambda}^{y(+)} \simeq \ii\osigma_{\lambda}$
are also obtained approximately.
Even if $|\ZZ|$ is not negligible, 
the atomic pump $\ocu$ multiplied by Eq.~\eqref{eq:<xy>+<yx>}
shows the even-order harmonics.
Further, for large enough $N \gg 1$,
the deviation from the unity can be neglected
in the similar way as the products of the operators are factorized
under the mean-field approximation in the macroscopic laser equation.

\section{Oscillating steady states in the Coulomb gauge} \label{eq:steady_Coulomb}
We decompose the five variables to frequency components as
$\AA(t) = \aa_{1}\ee^{-\ii\varOmega t} + \aa_{3}\ee^{-\ii3\varOmega t} + \cc$
($\BB$ to $\bb_n$, $\XX$ to $\xx_n$, and $\YY$ to $\yy_n$),
and $\ZZ(t) = \ZZ_0 + ( \zz_2\ee^{-\ii2\varOmega t} + \cc )$.
Then, neglecting highly oscillating terms,
the equations of the frequency components are obtained
in the Coulomb gauge as
\begin{subequations}
\begin{align}
\ddt\aaC_n
& = \ii n\varOmega \aaC_n - \wc \bbC_n, \\
\ddt\bbC_n
& = (\wc+4\rabi^2\wa-\ii\kappa) \aaC_n + \ii n\varOmega\bbC_n + 4\rabi\wa\yyC_n, \\
\ddt\xxC_1
& = (\ii\varOmega-\gamma_x)\xxC_1 - \wa \yyC_1
  + 2\rabi\wa\left( \ZZC_0\aaC_1 + \zzC_2^*\aaC_3 + \aaC_1^*\zzC_2 \right), \\
\ddt\xxC_3
& = (\ii3\varOmega-\gamma_x)\xxC_3 - \wa \yyC_3
  + 2\rabi\wa\left( \ZZC_0\aaC_3 + \aaC_1\zzC_2 \right), \\
\ddt\yyC_n
& = \wa \xxC_n + (\ii n\varOmega -\gamma_y) \yyC_n, \\
\ddt\ZZC_0
& = - \gamma_z\left(\ZZC_0 - \ocu\right)
  - 2\rabi\wa\left( \aaC_1^*\xxC_1 + \aaC_3^*\xxC_3 + \cc \right), \\
\ddt\zzC_2
& = (\ii2\varOmega- \gamma_z)\zzC_2
  - 2\rabi\wa\left( \aaC_1^*\xxC_3 + \aaC_1\xxC_1 + \xxC_1^*\aaC_3 \right).
\end{align}
\end{subequations}
In oscillating steady states, all of these derivatives should be zero.
Equations to be satisfied are finally reduced to
\begin{subequations} \label{eq:solved_Coulomb} 
\begin{align}
0 & = \left\{
    \frac{\varDelta_{c,1}\varDelta_{a,1}}{8\rabi^2\wc\wa^3} + \ocu
  + \left[ C_{|1|^21} + C_{|3|^21} |\eta|^2 + C_{(-1)^23}\eta \right] |\aaC_1|^2
  \right\} \aaC_1, \\
0 & = \left\{
    \frac{\varDelta_{c,3}\varDelta_{a,3}}{8\rabi^2\wc\wa^3} + \ocu
  + \left[ C_{|1|^23} + C_{|3|^23}|\eta|^2 + C_{1^3} \eta^{-1} \right] |\aaC_1|^2
  \right\} \eta\aaC_1 \ee^{\ii2\theta},
\end{align}
\end{subequations}
where $\theta$ is the phase of $\aaC_1 = |\aaC_1|\ee^{\ii\theta}$.
Unknown variables are $\varOmega$, $|\aaC_1| \in \mathbb{R}$ and a complex value
\begin{equation}
\eta = \frac{\aaC_3}{\aaC_1} \ee^{-\ii2\theta} \in \mathbb{C}.
\end{equation}
The other quantities in Eqs.~\eqref{eq:solved_Coulomb} are defined as follows
\begin{subequations}
\begin{align}
\varDelta_{c,n} & = \wc(\wc+4\rabi^2\wa) - n^2 \varOmega^2 - \ii\kappa\wc, \\
\varDelta_{a,n} & = \wa{}^2 + (\ii n\varOmega-\gamma_x)(\ii n\varOmega-\gamma_y),
\end{align}
\end{subequations}
\begin{subequations}
\begin{align}
C_{|1|^21}
& = - \frac{\Re[(\ii\varOmega-\gamma_y)\varDelta_{c,1}]}{\gamma_z\wc\wa}
  + \frac{(\ii\varOmega-\gamma_y)\varDelta_{c,1}}{2\wc\wa(\ii2\varOmega-\gamma_z)}, \\
C_{|3|^21}
& = - \frac{\Re[(\ii3\varOmega-\gamma_y)\varDelta_{c,3}]}{\gamma_z\wc\wa}
  + \frac{(\ii3\varOmega+\gamma_y)\varDelta_{c,3}^* - (\ii\varOmega-\gamma_y)\varDelta_{c,1}}{2\wc\wa(\ii2\varOmega+\gamma_z)}, \\
C_{(-1)^23}
& = \frac{(\ii3\varOmega-\gamma_y)\varDelta_{c,3} - (\ii\varOmega+\gamma_y)\varDelta_{c,1}^*}{2\wc\wa(\ii2\varOmega-\gamma_z)}
  + \frac{(\ii\varOmega+\gamma_y)\varDelta_{c,1}^*}{2\wc\wa(\ii2\varOmega+\gamma_z)}, \\
C_{|1|^23}
& = - \frac{\Re[(\ii\varOmega-\gamma_y)\varDelta_{c,1}]}{\gamma_z\wc\wa}
  + \frac{(\ii3\varOmega-\gamma_y)\varDelta_{c,3} - (\ii\varOmega+\gamma_y)\varDelta_{c,1}^*}{2\wc\wa(\ii2\varOmega-\gamma_z)}, \\
C_{|3|^23}
& = - \frac{\Re[(\ii3\varOmega-\gamma_y)\varDelta_{c,3}]}{\gamma_z\wc\wa}, \\
C_{1^3}
& = \frac{(\ii\varOmega-\gamma_y)\varDelta_{c,1}}{2\wc\wa(\ii2\varOmega-\gamma_z)}.
\end{align}
\end{subequations}
Once we get a solution of Eqs.~\eqref{eq:solved_Coulomb},
the other frequency components are obtained as
\begin{subequations}
\begin{align}
\aaC_3 & = \eta\aaC_1 \ee^{\ii2\theta}, \\
\bbC_n & = \frac{\ii n\varOmega}{\wc}\aaC_n, \\
\xxC_n & = \frac{(\ii n\varOmega-\gamma_y)\varDelta_{c,n}}{4\rabi\wc\wa^2}\aaC_n, \\
\yyC_n & = - \frac{\wa}{\ii n\varOmega-\gamma_y}\xxC_n, \\
\ZZC_0
& = \ocu - \frac{2\rabi\wa}{\gamma_z}
    \left( \aaC_1^*\xxC_1 + \aaC_3^*\xxC_3 + \cc \right), \\
\zzC_2
& = \frac{2\rabi\wa}{(\ii2\varOmega- \gamma_z)}
    \left( \aaC_1^*\xxC_3 + \aaC_1\xxC_1 + \xxC_1^*\aaC_3 \right),
\end{align}
\end{subequations}
while the phase $\theta$ of $\aaC_1$ can be chosen arbitrarily.

\section{Oscillating steady states in the electric-dipole gauge} \label{eq:steady_dipole}
In the same manner as in the Coulomb gauge,
the equations of the frequency components are obtained
in the electric-dipole gauge as
\begin{subequations}
\begin{align}
\ddt\aaP_n
& = \ii n\varOmega \aaP_n - \wc \bbP_n + 4\rabi\wc\xxP_n, \\
\ddt\bbP_n
& = (\wc-\ii\kappa) \aaP_n + \ii n\varOmega\bbP_n, \\
\ddt\xxP_n
& = (\ii n\varOmega -\gamma_x) \xxP_n - \wa \yyP_n, \\
\ddt\yyP_1
& = \wa \xxP_1 + (\ii\varOmega-\gamma_y)\yyP_1
  - 8\rabi^2\wc\left( \ZZP_0\xxP_1 + \zzP_2^*\xxP_3 + \xxP_1^*\zzP_2 \right)
  + 2\rabi\wc\left( \ZZP_0\bbP_1 + \zzP_2^*\bbP_3 + \bbP_1^*\zzP_2 \right), \\
\ddt\yyP_3
& = \wa \xxP_3 + (\ii3\varOmega-\gamma_y)\yyP_3
  - 8\rabi^2\wc\left( \ZZP_0\xxP_3 + \xxP_1\zzP_2 \right)
  + 2\rabi\wc\left( \ZZP_0\bbP_3 + \bbP_1\zzP_2 \right), \\
\ddt\ZZP_0
& = - \gamma_z\left( \ZZP_0 - \ocu \right)
  + 8\rabi^2\wc\left( \xxP_1^*\yyP_1 + \xxP_3^*\yyP_3 + \cc \right)
  - 2\rabi\wc\left( \bbP_1^*\yyP_1 + \bbP_3^*\yyP_3 + \cc \right), \\
\ddt\zzP_2
& = (\ii2\varOmega- \gamma_z)\zzP_2
  + 8\rabi^2\wc \left( \xxP_1^*\yyP_3 + \xxP_1\yyP_1 + \yyP_1^*\xxP_3 \right)
  - 2\rabi\wc \left( \bbP_1^*\yyP_3 + \bbP_1\yyP_1 + \yyP_1^*\bbP_3 \right).
\end{align}
\end{subequations}
The equations to be solved are
\begin{subequations} \label{eq:solved_PZW} 
\begin{align}
0 & = \left\{
    \frac{\varDelta_{c,1}\varDelta_{a,1}}{8\rabi^2\wc\wa\varOmega^2} + \ocu
  + \left[ C_{|1|^21} + C_{|3|^21} |\eta|^2 + C_{(-1)^23}\eta \right] |\aaP_1|^2
\right\}\aaP_1, \\
0 & = \left\{
    \frac{\varDelta_{c,3}\varDelta_{a,3}}{8\rabi^2\wc\wa\varOmega^2} + 9\ocu
  + \left[ C_{|1|^23} + C_{|3|^23}|\eta|^2 + C_{1^3} \eta^{-1} \right] |\aaP_1|^2
\right\} \eta \aaP_1 \ee^{\ii2\theta},
\end{align}
\end{subequations}
for unknown variables $\varOmega$, $|\aaP_1| \in \mathbb{R}$, and
$\eta = \ee^{-\ii2\theta}\aaP_3/\aaP_1 \in \mathbb{C}$.
The other quantities in Eqs.~\eqref{eq:solved_PZW} are defined as follows.
\begin{subequations}
\begin{align}
\varDelta_{c,n} & = \wc{}^2 - n^2 \varOmega^2 - \ii\kappa\wc, \\
\varDelta_{a,n} & = \wa{}^2 + (\ii n\varOmega-\gamma_x)(\ii n\varOmega-\gamma_y),
\end{align}
\end{subequations}
\begin{subequations}
\begin{align}
C_{|1|^21}
& = - \frac{\Re[(\ii\varOmega-\gamma_x)\varDelta_{c,1}]}{\gamma_z\wc\wa}
  + \frac{(\ii\varOmega-\gamma_x)\varDelta_{c,1}}{2\wc\wa(\ii2\varOmega-\gamma_z)}, \\
C_{|3|^21}
& = - \frac{\Re[(\ii3\varOmega-\gamma_x)\varDelta_{c,3}]}{\gamma_z\wc\wa}
  + \frac{(\ii3\varOmega+\gamma_x)\varDelta_{c,3}^* - 9(\ii\varOmega-\gamma_x)\varDelta_{c,1}}{2\wc\wa(\ii2\varOmega+\gamma_z)}, \\
C_{(-1)^23}
& = - \frac{(\ii3\varOmega-\gamma_x)\varDelta_{c,3} - 9(\ii\varOmega+\gamma_y)\varDelta_{c,1}^*}{6\wc\wa(\ii2\varOmega-\gamma_z)}
  - \frac{3(\ii\varOmega+\gamma_x)\varDelta_{c,1}^*}{2\wc\wa(\ii2\varOmega+\gamma_z)}, \\
C_{|1|^23}
& = - \frac{9\Re[(\ii\varOmega-\gamma_x)\varDelta_{c,1}]}{\gamma_z\wc\wa}
  + \frac{(\ii3\varOmega-\gamma_x)\varDelta_{c,3} - 9(\ii\varOmega+\gamma_x)\varDelta_{c,1}^*}{2\wc\wa(\ii2\varOmega-\gamma_z)}, \\
C_{|3|^23}
& = - \frac{9\Re[(\ii3\varOmega-\gamma_x)\varDelta_{c,3}]}{\gamma_z\wc\wa}, \\
C_{1^3}
& = - \frac{3(\ii\varOmega-\gamma_x)\varDelta_{c,1}}{2\wc\wa(\ii2\varOmega-\gamma_z)}.
\end{align}
\end{subequations}
The other frequency components are obtained as
\begin{subequations}
\begin{align}
\aaP_3 & = \eta\aaP_1 \ee^{\ii2\theta}, \\
\bbP_n & = - \frac{\wc-\ii\kappa}{\ii n\varOmega}\aaP_n, \\
\yyP_n & = - \frac{(\ii n\varOmega-\gamma_x)\varDelta_{c,n}}{\ii4 n\rabi\wc\wa\varOmega}\aaP_n, \\
\xxP_n & = \frac{\wa}{\ii n\varOmega-\gamma_x}\yyP_n, \\
\ZZP_0
& =  \ocu + \frac{\ii2\rabi\varOmega}{\gamma_z}
    \left( \aaP_1^*\yyP_1 + 3\aaP_3^*\yyP_3 - \cc \right), \\
\zzP_2
& = - \frac{\ii2\rabi\varOmega}{\ii2\varOmega- \gamma_z}
    \left( \aaP_1^*\yyP_3 - \aaP_1\yyP_1 - 3\yyP_1^*\aaP_3 \right).
\end{align}
\end{subequations}

\section{Other numerical results} \label{eq:numerical_dipole}
\begin{figure}[p]
\includegraphics[width=.8\linewidth]{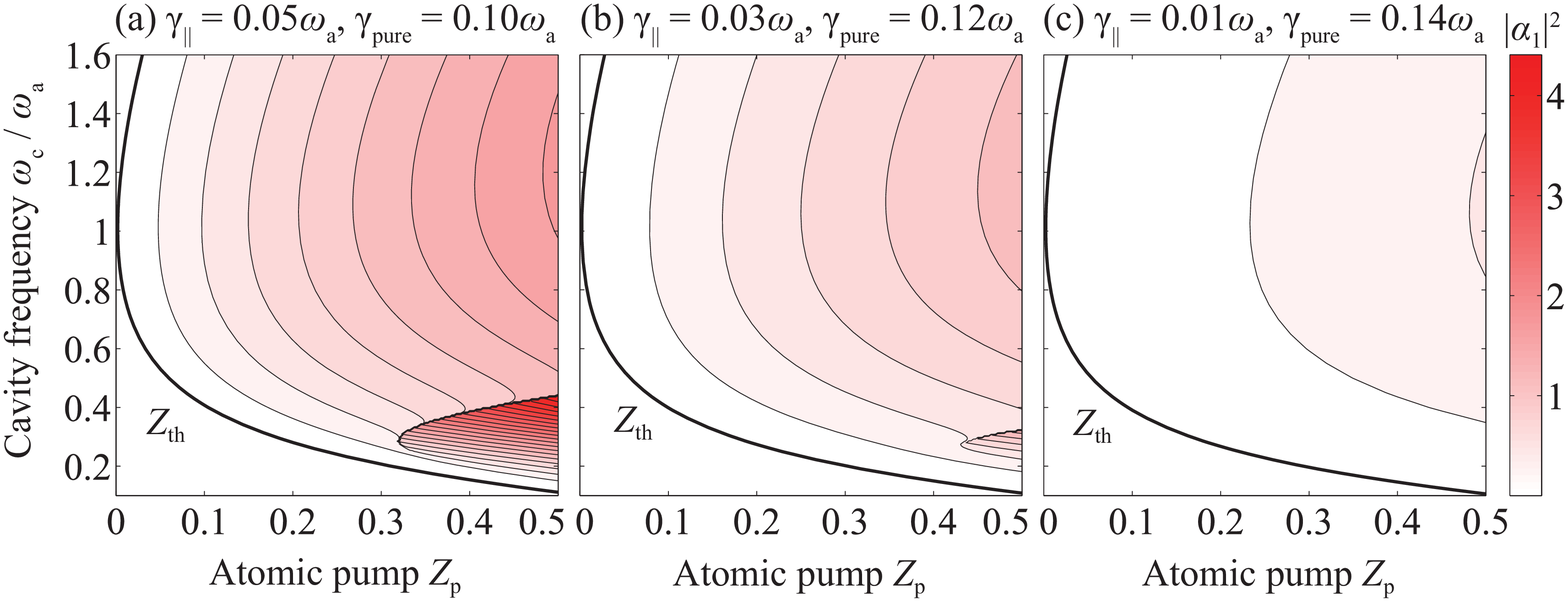}
\caption{Maps of ``laser'' solutions calculated in the electric-dipole gauge
for $\rabia = 0.4$ and $\kappa = 0.01\wa$.
The atomic dissipation rates are shown above the figures.
The unconventional solutions appear for strong enough interaction
and low enough cavity loss even in the electric-dipole gauge,
while they disappear by increasing the ratio of the pure dephasing
$\gammapure$ with keeping the total dissipation rate $\gammatotal=\gammaL+\gammapure$.}
\label{fig:1s}
\end{figure}
\begin{figure}[p]
\includegraphics[width=.8\linewidth]{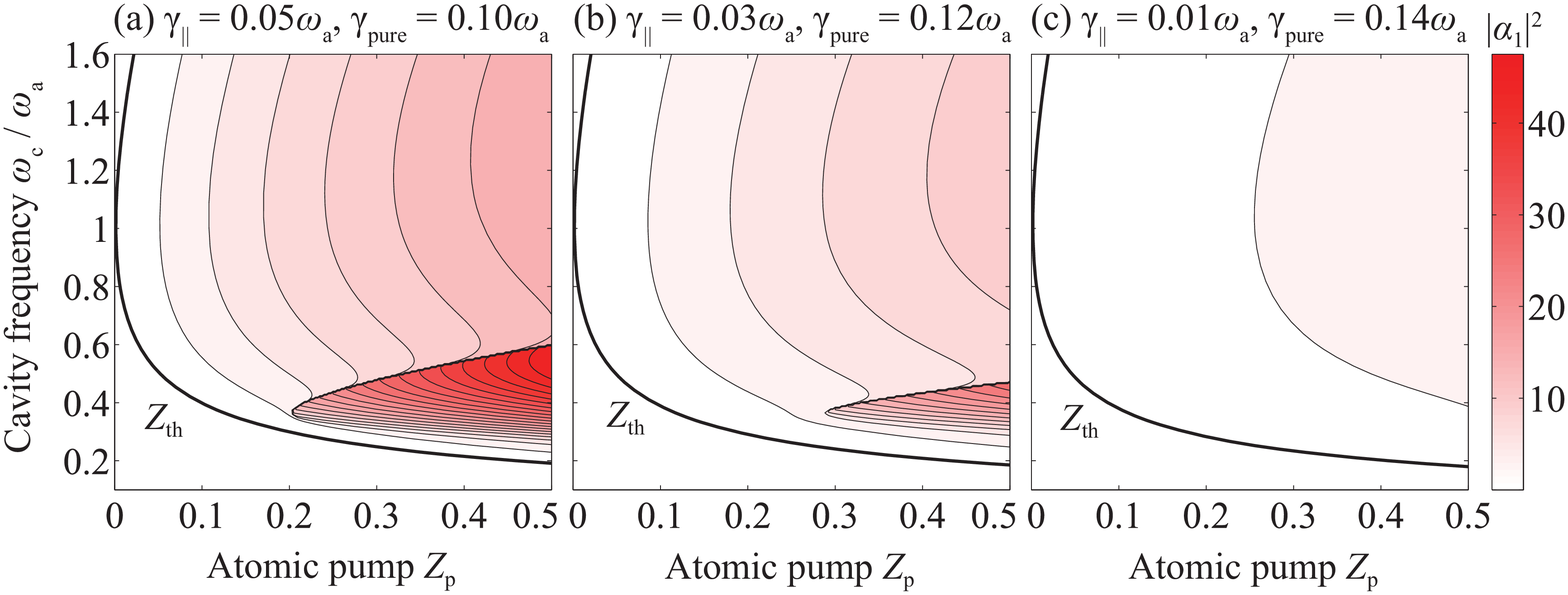}
\caption{Maps of ``laser'' solutions calculated in the electric-dipole gauge
for $\rabia = 0.15$ and $\kappa = 0.001\wa$.
The atomic dissipation rates are shown above the figures.
The unconventional solutions appear for strong enough interaction
and low enough cavity loss even in the electric-dipole gauge,
while they disappear by increasing the ratio of the pure dephasing
$\gammapure$ with keeping the total dissipation rate $\gammatotal=\gammaL+\gammapure$.}
\label{fig:2s}
\end{figure}
\begin{figure}[p]
\includegraphics[width=.8\linewidth]{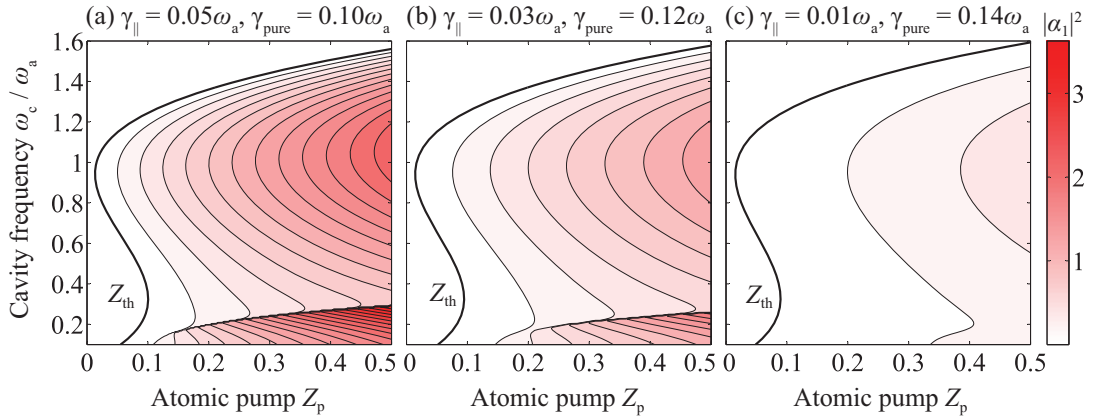}
\caption{Maps of ``laser'' solutions calculated in the Coulomb gauge
for $\rabia = 0.15$ and $\kappa = 0.01\wa$.
The atomic dissipation rates are shown above the figures
(Panel (a) is equivalent to Fig.~2(a) in the main text).
The unconventional solutions disappear by increasing the ratio of the pure dephasing
$\gammapure$ with keeping the total dissipation rate $\gammatotal=\gammaL+\gammapure$.}
\label{fig:3s}
\end{figure}
The maps of ``laser'' solutions in the electric-dipole gauge are plotted
in Fig.~\ref{fig:1s}(a)
for stronger electromagnetic interaction with matter $\rabia = 0.4$
and in Fig.~\ref{fig:2s}(a) for lower cavity loss $\kappa = 0.001\wa$
compared with the parameters in the main text.
Under these conditions, the bistability appears even in the electric-dipole gauge.

In Figs.~\ref{fig:1s}-\ref{fig:3s},
the dependence on the pure dephasing rate $\gammapure$ is also shown.
Figs.~\ref{fig:1s} and \ref{fig:2s} are calculated in the electric-dipole gauge,
and Figs.~\ref{fig:3s} is calculated in the Coulomb gauge
(Fig.~\ref{fig:3s}(a) is equivalent to Fig.~2(a) in the main text).
The pure-dephasing ratio $\gammapure/\gammatotal$ is changed with keeping
the total dissipation rates $\gammatotal = \gammaL + \gammapure = 0.15\wa$.
In the conventional laser theory \cite{gardiner04},
the laser occurs under the following condition (determing the threshold $\Zth$)
\begin{equation} \label{eq:cond_conv_laser} 
\frac{2\rabi^2\wa^2\ocu}{\kappa\gammatotal - (\wc-\varOmega)(\wa-\varOmega)} > 1,
\end{equation}
where the oscillation frequency is obtained as
\begin{equation}
\varOmega = \frac{\kappa\wa+\gammatotal\wc}{\kappa+\gammatotal}.
\end{equation}
In this way, the $\wc$-range showing the conventional laser does not depends
on the pure-dephasing ratio $\gammapure/\gammatotal$,
and this tendency basically survives even in Figs.~\ref{fig:1s}-\ref{fig:3s}.
Eq.~\eqref{eq:cond_conv_laser} is rewritten as
\begin{equation} \label{eq:cond2_conv_laser} 
2\rabi^2\wa^2\ocu
> \kappa\gammatotal\left[ 1 + \left(\frac{\wa-\wc}{\gammatotal+\kappa}\right)^2 \right].
\end{equation}
This relation basically determines the $\wc$-range of the ``laser''.
Then, since we suppose $\kappa \ll \gammatotal$,
the $\wc$-range is enlarged by lowering $\kappa$
and by heightening $\gammatotal$ and $\rabia$.

In either calculation in Figs.~\ref{fig:1s}-\ref{fig:3s},
the bistable ``laser'' solutions are obtained for $\wc \sim \wa/3$,
which is the requirement for the bistability as discussed in the main text.
However, by increasing the pure-dephasing ratio $\gammapure/\gammatotal$,
the unconventional solutions gradually disappear.
This is because the amplitudes of the electromagnetic fields
are diminished by increasing $\gammapure/\gammatotal$
as clearly seen in the figures.
Especially, the third harmonic amplitudes are diminished
more drastically (not shown in the figures),
because they appear as similar as the nonlinear optical effect.
For much lower $\gammapure/\gammatotal$,
the bistability appears more clearly (not shown in the figures).
Then, for obtaining the bistability,
we should prepare low enough $\kappa$ and $\gammapure$
and high enough $\gammaL$ and $\rabia$.

\begin{figure}[tbp]
\includegraphics[width=.6\linewidth]{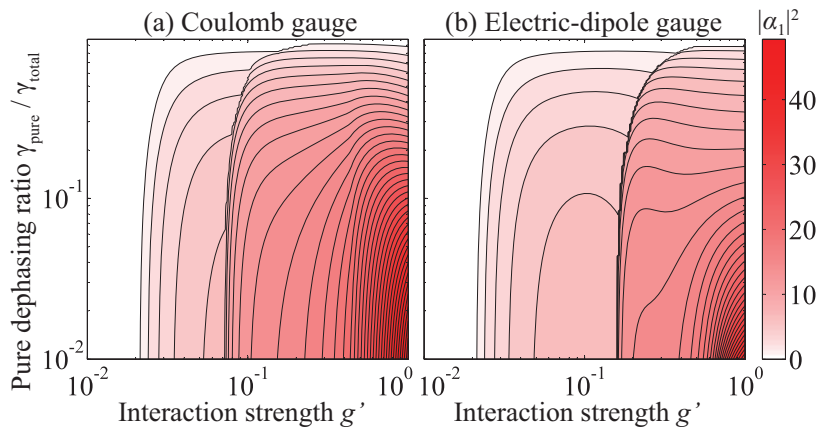}
\caption{Maps of ``laser'' solutions 
in (a) Coulomb gauge and (b) electric-dipole gauge.
The color indicates the maximum $|\aa_1|^2$ searched by changing $\wc$.
The larger value is chosen in the bistable case.
The maximum $|\aa_1|^2$ is plotted versus the interaction strength $\rabia$
and the pure-dephasing ratio $\gammapure/\gammatotal$
with keeping $\ocu = 0.5$, $\gammatotal = 0.15\wa$, and $\kappa = 0.01\wa$.}
\label{fig:4s}
\end{figure}
In Figs.~\ref{fig:4s},
we plot the maximum $|\aa_1|^2$ searched by changing $\wc$ in the two gauges.
The larger value is chosen in the bistable case,
and it is plotted versus the interaction strength $\rabia$
and the pure dephasing ratio $\gammapure/\gammatotal$
with keeping $\ocu = 0.5$, $\gammatotal = 0.15\wa$, and $\kappa = 0.01\wa$.
The relatively large $|\aa_1|^2$ for strong enough $\rabia$ indicates
the existence of the bistability.
In either gauge,
whereas the bistability is obtained clearly for low enough $\gammapure/\gammatotal$,
the visibility is lowered with the increase in $\gammapure/\gammatotal$.
This tendency indicates the disappearance of the bistability
demonstrated in Figs.~\ref{fig:1s}-\ref{fig:3s}.
Whereas the bistability is obtained
even for the relatively high $\gammapure/\gammatotal \sim 0.8$
as shown in Figs.~\ref{fig:1s}-\ref{fig:3s},
stronger $\rabia$ is required for the bistability as seen in Figs.~\ref{fig:4s}.
The bistable region is not largely enhanced
even in the limit of $\gammapure/\gammatotal \rightarrow 0$.
This is because
the pure dephasing ratio $\gammapure/\gammatotal$
basically does not change the parameter region showing $\varOmega \sim \wa/3$,
that is determined by Eq.~\eqref{eq:cond2_conv_laser}.
\end{widetext}


\end{document}